\documentclass[prb,twocolumn,showpacs,superscriptaddress,preprintnumbers,floatfix,amsmath,amssymb]{revtex4-1}

\usepackage{amssymb}
\usepackage{graphicx}
\usepackage{dcolumn}
\usepackage{bm}
\usepackage{color}
\hfuzz=\maxdimen
\begin{document}

\title{Physical properties and electronic structure of Sr$_2$Cr$_3$As$_2$O$_2$ containing CrO$_2$ and Cr$_2$As$_2$ square-planar lattices}

\author{Hao Jiang}
\affiliation{Department of Physics, Zhejiang University, Hangzhou
310027, China}

\author{Jin-Ke Bao}
\affiliation{Department of Physics, Zhejiang University, Hangzhou
310027, China}

\author{Hui-Fei Zhai}
\affiliation{Department of Physics, Zhejiang University, Hangzhou
310027, China}

\author{Zhang-Tu Tang}
\affiliation{Department of Physics, Zhejiang University, Hangzhou
310027, China}

\author{Yun-Lei Sun}
\affiliation{Department of Physics, Zhejiang University, Hangzhou
310027, China}

\author{Yi Liu}
\affiliation{Department of Physics, Zhejiang University, Hangzhou
310027, China}

\author{Zhi-Cheng Wang}
\affiliation{Department of Physics, Zhejiang University, Hangzhou
310027, China}

\author{Hua Bai}
\affiliation{Department of Physics, Zhejiang University, Hangzhou
310027, China}

\author{Zhu-An Xu}
\affiliation{Department of Physics, Zhejiang University, Hangzhou
310027, China} \affiliation{State Key Lab of Silicon Materials,
Zhejiang University, Hangzhou 310027, China}
\affiliation{Collaborative Innovation Centre of Advanced Microstructures, Nanjing 210093, China}

\author{Guang-Han Cao} \email[Correspondence should be sent to: ]{ghcao@zju.edu.cn}
\affiliation{Department of Physics, Zhejiang University, Hangzhou
310027, China} \affiliation{State Key Lab of Silicon Materials,
Zhejiang University, Hangzhou 310027, China} \affiliation{Collaborative Innovation Centre of Advanced Microstructures, Nanjing 210093, China}

\date{\today}

\begin{abstract}
We report the physical properties and electronic structure calculations of a layered chromium oxypnictide, Sr$_2$Cr$_3$As$_2$O$_2$, which crystallizes in a Sr$_2$Mn$_3$As$_2$O$_2$-type structure containing both CrO$_2$ planes and Cr$_2$As$_2$ layers. The newly synthesized material exhibits a metallic conduction with a dominant electron-magnon scattering. Magnetic and specific-heat measurements indicate at least two intrinsic magnetic transitions below room temperature. One is an antiferromagnetic transition at 291 K, probably associated with a spin ordering in the Cr$_2$As$_2$ layers. Another transition is broad, occurring at around 38 K, and possibly due to a short-range spin order in the CrO$_2$ planes. Our first-principles calculations indicate predominant two-dimensional antiferromagnetic exchange couplings, and suggest a KG-type (i.e. K$_2$NiF$_4$ type for CrO$_2$ planes and G type for Cr$_2$As$_2$ layers) magnetic structure, with reduced moments for both Cr sublattices. The corresponding electronic states near the Fermi energy are mostly contributed from Cr-3$d$ orbitals which weakly (modestly) hybridize with the O-2$p$ (As-4$p$) orbitals in the CrO$_2$ (Cr$_2$As$_2$) layers. The bare bandstructure density of states at the Fermi level is only $\sim$1/4 of the experimental value derived from the low-temperature specific-heat data, consistent with the remarkable electron-magnon coupling. The title compound is argued to be a possible candidate to host superconductivity.
\end{abstract}

\maketitle

\section{\label{sec:level1}Introduction}
The discovery of high-temperature superconductivity in cuprates\cite{bednorz,wmk} and iron pnictides\cite{hosono,cxh} represents two major advances in exploration of new superconducting materials in recent decades. Structurally, all the superconducting cuprates contain CuO$_2$ planes, while the iron-based superconductors possess Fe$_2$$X_2$ ($X$=As or Se) layers. Thus it is rational to explore superconductivity in a compound with similar two-dimensional (2D) networks containing 3$d$-transition-metal (3$d$TM) elements. Particularly, the two 2D networks structurally match in symmetry and size, and therefore it is possible to form a hybrid structure with both CuO$_2$ planes and Fe$_2$$X_2$ layers\cite{wang,ricci,jh}. However, previous efforts to synthesize the target materials were unsuccessful\cite{johnston10}, which is possibly because Cu (Fe) more easily bonds with $X$ (O) according to the ``Hard and Soft Acids and Bases'' concept\cite{hsab,jh}.

In fact, a similar compound, Sr$_2$Mn$_3$As$_2$O$_2$ (2322), which contains alternating MnO$_2$ sheets and Mn$_2$As$_2$ layers, was reported earlier in 1979\cite{Mn2322-79}. Later in 1990s, a series of 2322-type materials were synthesized and characterized\cite{Mn2322P-91,Zn2322,Mn2322A,Mn2322B,zwj,zwj2,cm98,kishio}, presumably due to the structural similarity with cuprate superconductors. Since the birth of high-temperature iron-based superconductors in 2008, there has been a renewed interest on studying the related systems\cite{jh,ozawa,clarke1,clarke2,johnston10,Mn2322Sb,Cr21222}. Among them, the Cr-containing iron arsenide Sr$_2$CrO$_2$Fe$_2$As$_2$\cite{Cr21222} seems to be very close to superconductivity, because it comprises of the superconductively active Fe$_2$As$_2$ layers and, a self doping may be realized through charge transfer between the distinct layers, like the cases in Sr$_2$VFeAsO$_3$\cite{cao} and Ba$_2$Ti$_2$Fe$_2$As$_4$O\cite{syl}. The absence of both superconductivity and spin-density wave in Sr$_2$CrO$_2$Fe$_2$As$_2$ is probably due to the mixed occupation of Cr (like the case in Sr$_4$Cr$_2$O$_6$Fe$_2$As$_2$\cite{tegel1,tegel2}), because the Cr doping kills superconductivity in iron pnictides\cite{sefat-Cr}. Similar mixed site occupancy was reported in Sr$_2$$M_3$As$_2$O$_2$ ($M_3$=Mn$_3$, Mn$_2$Cu, and MnZn$_2$)\cite{johnston10}. Owing to this problem, it is highly preferable to synthesize an isostructural material containing a single 3$d$TM element.

Recent observations of superconductivity in CrAs (under high pressures)\cite{ljl,kotegawa} and $A_2$Cr$_3$As$_3$ ($A$=K, Rb, Cs) under ambient pressure\cite{bao,Rb233,Cs233} motivate us to explore possible superconductivity in other Cr-based compounds. Materials containing Cr$_2$As$_2$ layers, isostructural to the Fe$_2$$X_2$ layers mentioned above, could be the possible candidates. So far, a few compounds contain Cr$_2$As$_2$ layers have been reported. BaCr$_2$As$_2$ and SrCr$_2$As$_2$ were synthesized in 1980\cite{Ba122Cr}, but it was after the discovery of superconductivity in doped BaFe$_2$As$_2$\cite{johrendt} that the physical properties and electronic structure of BaCr$_2$As$_2$ were investigated\cite{singh}. The results show that BaCr$_2$As$_2$ is an itinerant antiferromagnetic metal, with a possible G-type magnetic order in the ground state. Very recently, similar properties for the Cr electronic states were reported in an analogous compound EuCr$_2$As$_2$\cite{hossain}. Park et al.\cite{park} synthesized a series of ZrCuSiAs-type oxypnictides $Ln$CrAsO ($Ln$=La, Ce, Pr, and Nd). All the members are metallic, and for LaCrAsO, a G-type antiferromagnetic (AFM) order with a moment of 1.57 $\mu_\mathrm{B}$ along the $c$ axis is revealed at 300 K by powder neutron diffractions. Up to now, superconductivity has not been observed in any of the Cr$_2$As$_2$-layer based materials, presumably because of the heavy $d-p$ hybridizations between Cr-3$d$ and As-4$p$ orbitals\cite{singh,hossain,park}.

In this work, we successfully synthesized a 2322-type Cr-based oxypnictide, Sr$_2$Cr$_3$As$_2$O$_2$, for the first time. The new material is found to be metallic with a dominant magnetic scattering. Three possible magnetic transitions associated with the Cr-spin orderings are observed. Our first-principles calculations suggest a KG-type magnetic ground state with reduced magnetic moments for the Cr sublattices. We also found that Cr-3$d$ orbitals dominate the electronic states near the Fermi energy. In contrast with the strong $d-p$ hybridizations in BaCr$_2$As$_2$\cite{singh}, EuCr$_2$As$_2$\cite{hossain} and LaCrAsO\cite{park}, the $d-p$ hybridizations in Sr$_2$Cr$_3$As$_2$O$_2$ are unexpectedly modest, resembling the scenarios in Fe$_2$As$_2$-layer based materials\cite{singh2}. This character, together with the structural and physical properties, implies the possibility to introduce superconductivity in the present system either through chemical dopings or via applying high pressures.

\section{\label{sec:level2}Experimental Methods}

\paragraph{Sample's synthesis} The Sr$_2$Cr$_3$As$_2$O$_2$ polycrystalline sample was synthesized by solid-state reactions in sealed evacuated quartz tubes. The starting materials were SrO powder (99.5\%), metal Cr powder (99.9\%) and As pieces (99.999\%). The SrO powder was heated at 1473 K for 24 h before using. We found that nearly single-phase samples could be prepared only by a small deviation of stoichiometry with 2$\sim$5\% oxygen deficiency. The desired mixtures were calcined in sealed evacuated quartz tubes in a muffle furnace at 1073 K for 24 h. The reacted mixtures were then ground for homogenization in an agate mortar, pressed into pellets, and sintered at 1223 K for 72 h. Note that an argon-filled glove box was employed for the operations above to avoid the possible reactions with water and oxygen.

\paragraph{Crystal structure determination} Powder x-ray diffractions (XRD) were carried out on a PANalytical x-ray diffractometer with a monochromatic CuK$_{\alpha1}$ radiation. The lattice parameters were determined using Si powders as the internal standard material. The crystal structure was determined by a Rietveld refinement\cite{rietan-fp} by employing the 2322-type structure\cite{Mn2322B} as a starting point. The final convergent refinement gives a weighted reliable factor of $R_{\text{wp}}$=7.3\% and a goodness-of-fit $\chi^2$=3.0, indicating quite good reliability for the crystal structure refined.

\paragraph{Physical property measurements} All the samples measured are from the same sintered pellet of Sr$_2$Cr$_3$As$_2$O$_2$. A Quantum Design physical property measurement system (PPMS-9) was employed to measure the electrical resistivity and specific heat. In the resistivity measurement, four electrodes were made on a 2.0$\times$1.2$\times$0.9 mm$^3$ rectangular sample bar using fine gold wires and silver paste. The specific heat was measured via a relaxation method using a 30.3 mg sample plate. We also employed a Quantum Design magnetic property measurement system (MPMS-5) to measure the sample's magnetization as functions of temperature and magnetic field.

\paragraph{Electronic structure calculations} Electronic structure calculations were performed within the generalized gradient approximation\cite{GGA} employing the Vienna Ab-initio Simulation Package (VASP)\cite{vasp}. The refined experimental crystal structure was used for the calculations. The plane-wave basis energy cutoff was set at 540 eV. As for the G-type magnetic ground state with a $\sqrt{2}\times\sqrt{2}$ superlattice, a $12\times12\times4$ and $18\times18\times6$  \textbf{k}-mesh grided by the scheme of Monkhorst and Pack\cite{MP} was used for the static and density-of-states (DOS) calculations, respectively.

\section{\label{sec:level3}Results and discussion}

\subsection{\label{subsec:level1}Crystal structure}

\begin{figure}
\includegraphics[width=8cm]{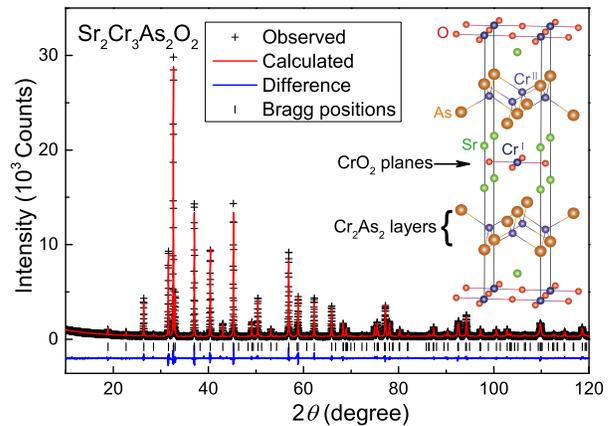}
\caption{\label{XRD}(Color online) Powder X-ray diffraction and its Rietveld refinement profile for Sr$_2$Cr$_3$As$_2$O$_2$. The inset shows the crystal structure which contains alternating CrO$_2$ planes and Cr$_2$As$_2$ layers.}
\end{figure}

Figure~\ref{XRD} shows the XRD pattern of the Sr$_2$Cr$_3$As$_2$O$_2$ powdered sample. Most of the reflections can be well indexed by a tetragonal unit cell whose size is close to that of Sr$_2$Mn$_3$As$_2$O$_2$. The unindexed reflections have their intensities lower than 2\%, which possibly comes from unreacted SrO, CrAs and/or Cr$_2$As. We were able to make a successful Rietveld refinement, which yields the crystallographic data listed in Table~\ref{parameter}. The unit-cell size is reasonably between those of Sr$_2$Mn$_3$As$_2$O$_2$\cite{Mn2322-79} and Sr$_2$CrO$_2$Fe$_2$As$_2$\cite{Cr21222}. Compared with Sr$_2$Mn$_3$As$_2$O$_2$\cite{Mn2322B}, the $a$ axis is 3.4\% smaller, while the $c$ axis is shortened merely by 0.1\%. If compared with Sr$_2$CrO$_2$Fe$_2$As$_2$\cite{Cr21222}, however, the $a$ axis is 0.3\% larger, and the $c$ axis expands by 2.2\%. The variations in lattice constants can be basically understood by the effective ionic radius\cite{shannon} of Cr$^{2+}$ (0.73 \r{A}) which is between those of Mn$^{2+}$ (0.83 \r{A}) and Fe$^{2+}$ (0.61 \r{A}).

\begin{table}
\caption{\label{parameter}
Crystallographic data of Sr$_2$Cr$_3$As$_2$O$_2$ with $a$ = 4.0079(1) \r{A}, $c$ = 18.8298(3) \r{A} and space group $I4/mmm$ (No. 139), refined from powder XRD data at room temperature.}
\begin{ruledtabular}
\begin{tabular}{cccccc}
Atom& Wyckoff& $x$ &$y$&$z$&$B (\mathrm{\r{A}^{-2}})$ \\
\hline
Sr &4$e$&0& 0  &0.4123(1)  &0.12(3)\\
Cr$^{\mathrm{I}}$& 2$a$&0& 0 &0  &0.11(6) \\
Cr$^{\mathrm{II}}$& 4$d$&0&0.5  &0.25  &0.08(4)\\
As& 4$e$&0&  0 &0.8299(1)  &0.10(3)\\
O& 4$c$&0&0.5   &0  &0.43(17)\\
\end{tabular}
\end{ruledtabular}
\end{table}

There are two distinct Cr sites, which is clearly illustrated in the inset of Fig.~\ref{XRD}. The 2$a$-site Cr$^{\mathrm{I}}$ is located within the CrO$_2$ planes, notably coordinated with two additional arsenic atoms at the apexes of an elongated octahedron. On the other hand, the 4$d$-site Cr$^{\mathrm{II}}$ bonds with four As stoms in the Cr$_2$As$_2$ layers. In the crystal chemistry point of view, the structure comprises of ``infinite layer" SrCrO$_2$ (analogous to the infinite-layer cuprate SrCuO$_2$) building blocks that are sandwiched by the ThCr$_2$Si$_2$-type (half a unit cell) SrCr$_2$As$_2$ blocks. Note that the ternary compounds ``SrCrO$_2$" have not been synthesized yet. This means that, the CrO$_2$ sheets, rarely with divalent Cr in oxides, are stabilized by the SrCr$_2$As$_2$ blocks.

The Cr$^{\mathrm{I}}$ valence can be evaluated by its bond valence sum (BVS)\cite{bvs} (note that BVS calculation for Cr$^{\mathrm{II}}$ does not make sense because of the substantial covalent Cr$-$As bonding). Using the formula $\sum$ exp$(\frac{R_{0}-d_{ij}}{0.37})$, where $R_{0}$ is 1.73 (2.34) \r{A} for a Cr$^{2+}-$O (Cr$^{2+}-$As) bond and $d_{ij}$ are the measured bond distances between Cr$^{\mathrm{I}}$ and the coordinating anions, we obtained a Cr$^{\mathrm{I}}$ BVS value of 2.10. This is close to the apparent Cr valence, but it may also suggest a substantial charge transfer between the two Cr sites, like the case in Ba$_2$Ti$_2$Fe$_2$As$_4$O\cite{syl}.

The As$-$Cr$^{\mathrm{II}}-$As bond angle may be an important parameter governing the next-nearest-neighbor (NNN) magnetic exchange interactions. In iron-based superconductors, for example, a regular tetrahedron of FeAs$_4$ with the diagonal As$-$Fe$-$As angle of 109.5$^{\circ}$ was found to be optimal for achieving the highest superconducting transition temperature\cite{zhaoj,lee}. Here in Sr$_2$Cr$_3$As$_2$O$_2$, the As$-$Cr$^{\mathrm{II}}-$As bond angle is 106.2$^{\circ}$, which indicates that
the CrAs$_4$ tetrahedron is elongated along the $c$ axis. Similar local structure distortions exist in other Cr-based materials such as $A_E$Cr$_2$As$_2$ ($A_E$=Ba, Sr, Eu)\cite{Ba122Cr,hossain} and LaCrAsO\cite{park}, which could be related to the Jahn-Teller instability for a Cr$^{2+}$ ion with 3$d^{4}$ electronic configuration. The crystal field of the distorted CrAs$_4$ tetrahedron leads to the highest energy level for the $d_{xy}$ orbital. Owing to the strong Hund's rule coupling, the four electrons occupy the other four 3$d$ orbitals, respectively, giving an $S$=2 local spin within a simplified ionic scenario. Here we note that the local structure distortion generally reduces the NNN coupling, which may serve as a possible reason for the G-type magnetic ground state instead of a striped AFM state in the iron-based materials.

\subsection{\label{subsec:level2}Electrical resistivity}

\begin{figure}
\includegraphics{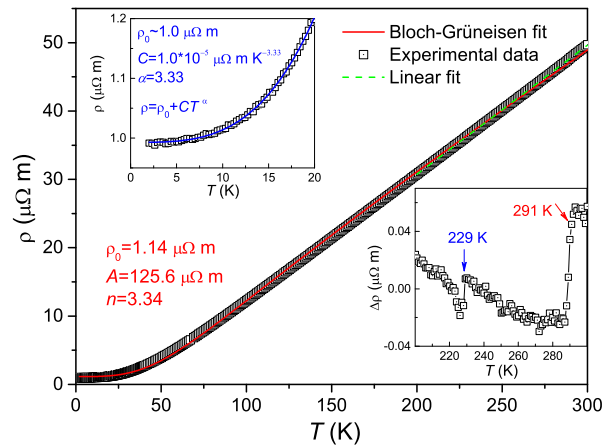}
\caption{\label{resistivity} (Color online) Temperature dependence of the electrical resistivity for the Sr$_2$Cr$_3$As$_2$O$_2$ polycrystalline pellets, which basically obeys an extended Bloch-Gr\"{u}neisen formula (see the text in details). The upper-left inset shows a power-law dependence in the low-temperature regime. The bottom-right inset plots the resistivity difference $\Delta \rho$ between the experimental data and the linearly fitted ones from 200 to 300 K.}
\end{figure}

The temperature dependence of resistivity, $\rho(T)$, for the Sr$_2$Cr$_3$As$_2$O$_2$ polycrystalline sample is shown in Fig.~\ref{resistivity}. The $\rho(T)$ curve is metallic with a large residual-resistivity ratio, RRR$\sim$50 (a ratio of the resistivity at room temperature and at 2 K here), suggesting that the grain boundaries convey minor contributions. Given the layered crystal structure, the $\rho(T)$ data may reflect the intrinsic \emph{in-plane} property. The $\rho(T)$ data in the full temperature range follow an extended Bloch-Gr\"{u}neisen formula with $n$ being a variable,
\begin{equation}
 \rho(T)=\rho_{0}+A \left(\frac{T}{\Theta_{R}}\right)^{n}\int_{0}^{\Theta_{R}/T}\frac{x^n}{(e^x-1)(1-e^{-x})}dx
\label{eq:one}.
\end{equation}
By adopting the Debye temperature ($\Theta_R$=320 K) from the specific-heat analysis below, the data fitting yields $n$=3.34. The $n$ value is fully consistent with the power $\alpha$=3.33, as fitted by the power-law function, $\rho(T)=\rho_0+CT^{\alpha}$, using the $\rho(T)$ data in the low-temperature regime (2 K $<T<$ 20 K). The yielded $n$ or $\alpha$ value implies a dominant electron-magnon scattering, rather than an ordinary electron-phonon scattering with $n$=5 or, an electron-electron interactions with $n$=2.

It is surprising that no appreciable anomaly can be detected from the resistivity data, although the following magnetic and specific-heat measurements indicate three possible magnetic transitions below room temperature. Furthermore, the magnetoresistance is vanishingly small from 2 to 300 K under a magnetic field of 50 kOe. Nevertheless, on closer examination of the data, one may see the corresponding responses. As shown in the bottom-right inset of Fig.~\ref{resistivity}, the subtraction of the linearly fitted data yields two resistivity drops at 291 K and 229 K, respectively. The magnitude of the larger drop is only 0.15\% of the absolute value of resistivity, suggesting a slight change in magnetic scattering. The possible reason is given below. Note that the $\rho(T)$ behavior quite resembles those of BaCr$_2$As$_2$\cite{singh} and LaCrAsO\cite{park} both of which contain the Cr$_2$As$_2$ layers. On the other hand, Sr$_2$CrO$_2$Fe$_2$As$_2$, which contains both Fe$_2$As$_2$ and CrO$_2$ planes, shows a semiconducting-like behavior\cite{Cr21222}. This indicates that the CrO$_2$ planes are not metallic. If this is true for Sr$_2$Cr$_3$As$_2$O$_2$, the observed metallic conduction predominately comes from the Cr$_2$As$_2$ layers. As discussed below, the Cr spins in the Cr$_2$As$_2$ layers are 2D antiferromagnetically correlated far above room temperature, and change in magnetic scattering at the transition temperature could be very limited, as being observed.

\subsection{\label{subsec:level3}Magnetic properties}

\begin{figure}
\includegraphics{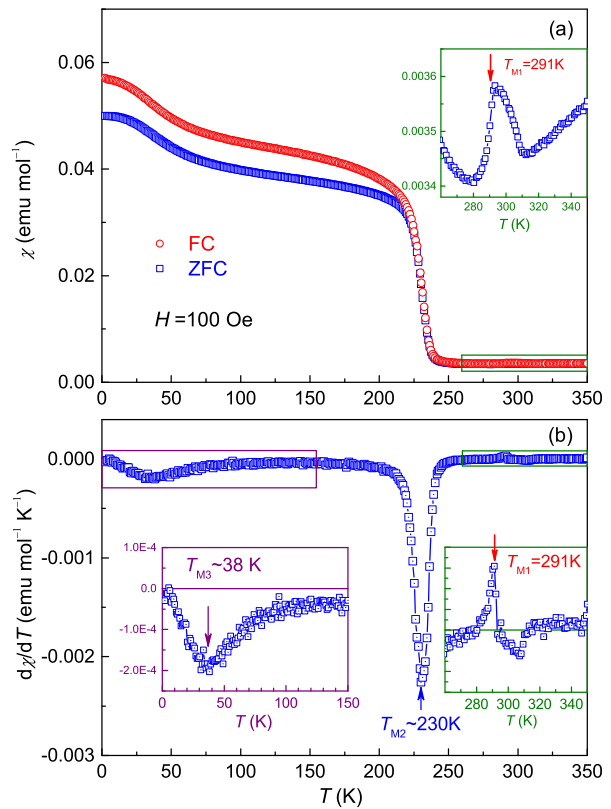}
\caption{Temperature dependence of magnetic susceptibility (a) and its derivative (b) for the Sr$_2$Cr$_3$As$_2$O$_2$ sample. The insets zoom in on areas of magnetic anomalies around $T_{\mathrm{M1}}$ and $T_{\mathrm{M3}}$. \label{mt}}
\end{figure}

\begin{figure}
\includegraphics[width=7cm]{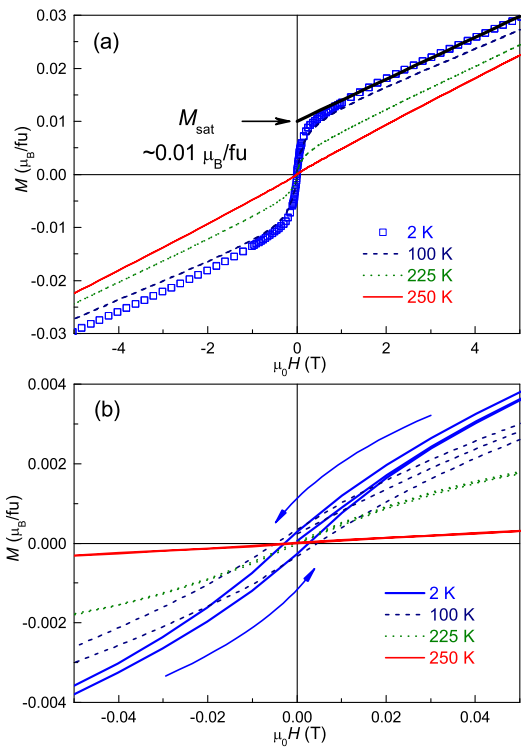}
\caption{Field dependence of magnetization at fixed temperatures for Sr$_2$Cr$_3$As$_2$O$_2$. The bottom panel shows an expanded plot from which a magnetic hysteresis can be seen at low temperatures.\label{mh}}
\end{figure}

Figure~\ref{mt} shows the temperature dependence of magnetic susceptibility ($\chi$) and its derivative for the Sr$_2$Cr$_3$As$_2$O$_2$ sample. First, a ferromagnetic/ferrimagnetic like transition can be seen at $T_{\mathrm{M2}}\sim$230 K. Bifurcation of $\chi$ can be observed in zero-field-cooling (ZFC) and field-cooling (FC) modes at 100 Oe, which is confirmed by the $M(H)$ curve shown in Fig.~\ref{mh}(b). However, the `net' spontaneous magnetization is very low. At 2 K (far below $T_{\mathrm{M2}}$), the extrapolated `saturation' magnetization is only 0.11 emu g$^{-1}$, or equivalently, 0.01 $\mu_{\mathrm{B}}$/fu (fu denotes formula unit). This small ordered moment usually suggests the existence of ferromagnetic impurities, which likely conforms to the vanishingly small anomaly in the $\rho(T)$ data above. We note that the ``Cr$_3$As$_2$'' and ``Cr$_4$As$_3$'' phases were earlier reported to be ferromagnetic (FM) with a Curie temperature from 213 to 257 K\cite{hn,yuzuri}, though these impurity phases could not be clearly identified by our XRD experiments. Furthermore, the saturation magnetization value was found to be sample dependent, which also supports the extrinsic nature of the weak FM transition.

Nevertheless, other two magnetic transitions, shown the insets of Fig.~\ref{mt}, are probably intrinsic. The first one occurs at $T_{\mathrm{M1}}$=291 K, accompanied with a drop in $\chi$, suggestive of an AFM transition. The magnetic transition is supported by the following specific-heat data which indicate a second-order-like phase transition at the same temperature. Above 310 K, $\chi$ increases almost linearly with temperature, reminiscent of the magnetic behavior in the parent compounds of iron-based superconductors\cite{johnston}. The latter is explained in terms of itinerant-electron AFM spin fluctuations\cite{zgm}, or alternatively, with an usual 2D short-range AFM spin correlations. The minimum in $\chi$ at 310 K may be due to superposition of a linear-like term and a Curie-Weiss term, $\chi_{\mathrm{CW}}=C/(T-\theta)$. The latter probably comes from the local-moment paramagnetism in CrO$_2$ planes, as is seen in Sr$_4$Cr$_2$O$_6$Fe$_2$As$_2$\cite{tegel1} and Sr$_2$CrO$_2$Fe$_2$As$_2$\cite{Cr21222}.

An additional magnetic anomaly takes place around $T_{\mathrm{M3}}$=38 K, where d$\chi$/d$T$ shows a broad valley. As is known, $\chi(T)$ mostly shows a Curie-Weiss `tail' for $T\rightarrow$ 0 K because of inevitable paramagnetic impurities and/or defects. Therefore, the observed saturation in $\chi$ at low temperatures here suggests that the intrinsic $\chi$ has to go down with $T\rightarrow$ 0 K. Namely, a broad peak at $T_{\mathrm{M3}}$ is expected for the intrinsic $\chi(T)$, similar to the observation in Sr$_2$Mn$_3$As$_2$O$_2$\cite{Mn2322A}, where 2D AFM correlations in the MnO$_2$ planes were revealed by neutron diffractions\cite{Mn2322B}. Below we will see that there is a corresponding specific-heat anomaly in the same temperature regime.

\subsection{\label{subsec:level4}Specific heat}

\begin{figure}
\includegraphics[width=7cm]{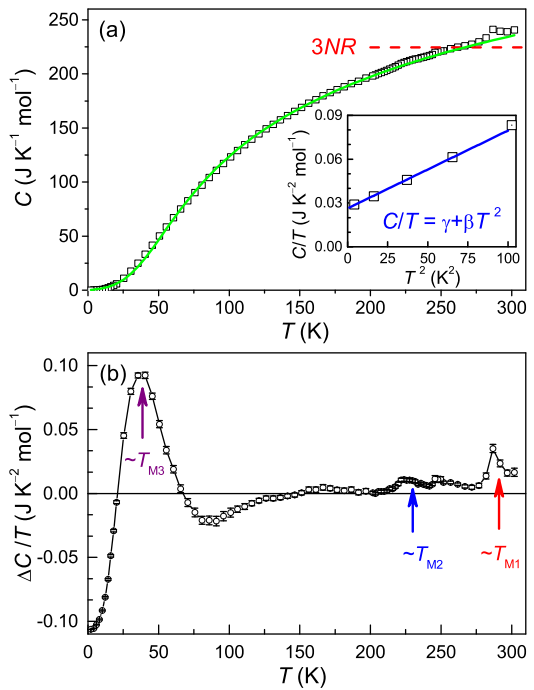}
\caption{\label{specific-heat}(Color online) Temperature dependence of specific heat for Sr$_2$Cr$_3$As$_2$O$_2$. The upper inset shows the low-temperature specific-heat data, plotted with $C/T$ vs $T^2$. The bottom panel plots the specific-heat difference between the experimental data and the fitted ones, from which three possible phase transitions can be identified.}
\end{figure}

Figure~\ref{specific-heat} shows the temperature dependence of specific heat, $C(T)$, for the Sr$_2$Cr$_3$As$_2$O$_2$ sample. The room-temperature specific-heat value exceeds the Dulong-Petit limit, 3$NR$ ($N$ is the number of atoms in a formula unit, and $R$ is the gas constant), by 7.1\%. This suggests additional contributions from electrons as well as from possible magnetic excitations. Subtle anomalies can be seen around 230 and 290 K, which seem to be related to the magnetic transitions observed above.

To extract these anomalies, we fit the $C(T)$ data by employing a combined formula with both Einstein and Debye components (represent optic and acoustic phonons, respectively)\cite{einstein}, in addition to a linear term $BT$ (to represent the electronic and magnon contributions, for simplicity\cite{linear}). Inspired by the successful treatment for LaB$_6$\cite{mandrus}, here we approximately treat the oxygen ions (with a fraction of 2/9) as independent Einstein oscillators embedded in a Debye framework composed of the heavier elements. Therefore, the fitting formula contains only three variable parameters, $\theta_{\mathrm{E}}$ (Einstein temperature), $\theta_{\mathrm{D}}$ (Debye temperature) and the coefficient $B$,
\begin{equation}
C(T)=\frac{2}{9}C_{\mathrm{E}}(\theta_{\mathrm{E}},T)+\frac{7}{9}C_{\mathrm{D}}(\theta_{\mathrm{D}},T)+BT
\label{eq:tot},
\end{equation}
where
\begin{equation}
C_{\mathrm{E}}(\theta_{\mathrm{E}},T)=3NR\left(\frac{\theta_{\mathrm{E}}/T}{e^{\theta_{\mathrm{E}}/T}-1}\right)^{2}e^{\theta_{\mathrm{E}}/T}
\label{eq:one},
\end{equation}
and
\begin{equation}
C_{\mathrm{D}}(\theta_{\mathrm{D}},T)=9NR\left(\frac{T}{\theta_{\mathrm{D}}}\right)^{3}\int_{0}^{\theta_{\mathrm{D}}/T}\frac{x^{4}e^x}{(e^{x}-1)^2}dx
\label{eq:two}.
\end{equation}
As seen in Fig.~\ref{specific-heat}(a), the data fitting from 2 to 200 K is quite successful, which yields $\theta_{\mathrm{E}}$=759.6 K, $\theta_{\mathrm{D}}$=323.4 K and $B$=0.134 J K$^{-2}$ mol$^{-1}$. The $\theta_{\mathrm{D}}$ value is consistent with the result from the low-temperature data analysis (see below). Meanwhile, the high $\theta_{\mathrm{E}}$ value is reasonable because of the small mass for oxygen atoms. The $B$ value is about five times of the electronic specific-heat coefficient from the analysis on the low-temperature data, which suggests substantial contributions from magnons.

Then, we may extract the anomalies related to the possible magnetic transitions by subtracting the fitted curve from the experimental data. The resulting $\Delta C$ may reflect the anomalies due to magnetic transitions. To estimate the magnetic entropy intuitively, $\Delta C$ is divided by $T$, which is plotted as a function of temperature in Fig.~\ref{specific-heat}(b). Three possible transitions can be identified, corresponding to the above magnetic anomalies. First, a specific-heat jump at 291 K coincides the AFM transition at the $T_{\mathrm{M1}}$ stated above. The magnetic entropy, defined by $S_{\mathrm{m}}=\int_{0}^{T}\frac{C_{\mathrm{m}}}{T}dT$, is estimated to be $S_{\mathrm{M1}}\sim$ 0.6 J K$^{-1}$ mol$^{-1}$, which is much lower than the expected value of $R$ln$(2S+1)$. This is consistent with the 2D AFM spin correlations inferred from the positive temperature coefficient of susceptibility above 310 K. Second, a minor broad anomaly can also be detected at $\sim$230 K, corresponding to the weak FM transition at $T_{\mathrm{M2}}$. The magnetic entropy is even lower: $S_{\mathrm{M2}}\sim$ 0.2 J K$^{-1}$ mol$^{-1}$. This seems to be reasonable since the ordered moment is merely 0.01 $\mu_{\mathrm{B}}$/fu which is likely to be an extrinsic property. Thirdly, there is a broad ``jump'' in $\Delta C/T$ around $T_{\mathrm{M3}}$, suggesting an additional magnetic ordering. The corresponding magnetic entropy is sizable with $S_{\mathrm{M3}}\sim$ 2.5 J K$^{-1}$ mol$^{-1}$. Note that the estimated entropy contains a considerably large uncertainty, because it is based on a subtraction where the subtrahend is from the data fitting. Nevertheless, the three magnetic anomalies are qualitatively indicated.

In the low-temperature regime, the lattice contribution of specific heat well obeys the Debye $T^3$ law, and the electronic contribution better satisfies the linear temperature dependence (because $T\ll T_{\mathrm{F}}$). Indeed, the $C/T$ vs $T^2$ relation is essentially linear in a low-temperature range from 2 to 10 K. The linear fit with the formula $C/T = \gamma + \beta T^2$ yields an intersect of $\gamma$ = 26.3 mJ K$^{-2}$ mol$^{-1}$ and a slope of $\beta$ = 0.532 mJ K$^{-4}$ mol$^{-1}$. Utilizing the formula $\theta_{\text{D}}=[(12/5)NR\pi^{4}/\beta]^{1/3}$, we obtain a Debye temperature of 320.3 K, which is in agreement with the fitted result with Eq. (2).

\subsection{\label{subsec:level5}Density-functional calculations}

To understand the physical properties above, we performed first-principles calculations based on density-functional theory. First, we calculated the magnetic energy ($E_{\mathrm{m}}$) for each of the possible magnetic states, defined by the energy difference with that of a non-spin-polarized (nonmagnetic) calculation. Because of two Cr sites in different layers, the possible spin configurations are the combination of the magnetic states in the two sublattices. For each Cr sublattice, the basic magnetic states are non-magnetism (N), ferromagnetism (F) and antiferromagnetism (AF). There are four simple AFM configurations, namely, i) A type (in-plane F and interplane AF), ii) C type (in-plane N\'{e}el AF and interplane F), iii) G type (in-plane N\'{e}el AF and interplane AF), and iv) S type (in-plane striped AF and interplane F\cite{s-type}). Note that the Cr$^\mathrm{I}$ sublattice is body centered, and for the in-plane N\'{e}el AF case, the classification into C and G types is not applicable. Here we consider the K$_2$NiF$_4$-type (abbreviated as K-type) spin order\cite{K2NiF4}. Thus, we may have 20 possible spin orders apart from other magnetic configurations with nonmagnetic states.

Table~\ref{ground states} lists the magnetic energies of selected magnetic states, from which we may catch the information about the magnetic ground state. Owing to the 2D structural characteristic, we search the stable spin order in 2D. In this way, C-, G- or K-type magnetism shares the same spin configuration in 2D. As can be seen in Table~\ref{ground states}, first of all, the magnetic state with both nonmagnetic Cr sublattices is unstable with respect to the magnetically ordered states. Secondly, under the assumption with nonmagnetic Cr$^\mathrm{I}$, the most stable magnetic configuration is Cr$^\mathrm{I}$N$-$Cr$^{\mathrm{II}}$C, whose $E_{\mathrm{m}}$ is 0.81 eV below that of the nonmagnetic state. The energy lowering actually comes from the Cr$^\mathrm{II}$ sublattice, suggesting strong magnetic exchange interactions in the Cr$_2$As$_2$ layers. Similarly, the energy gain for the magnetic order in the CrO$_2$ planes is about 1.1 eV. Thirdly, the magnetic states with the same planar spin configurations have nearly the same magnetic energy, indicating that the 2D planar magnetic interactions predominate. Overall, the planar N\'{e}el AF configurations (in C, G or K type) save the most energy for both Cr sublattices. Consequently, the most probable magnetic ground state is of K and G (hereafter abbreviated as KG) type for Cr$^{\mathrm{I}}$ and Cr$^{\mathrm{II}}$ sublattices, respectively. With the consideration of spin-orbit coupling, $E_{\mathrm{m}}$ is further reduced. The spin direction is found to be either along the crystallographic $c$ axis, or the $a$ axis.

\begin{table}[b]
\centering
\caption{\label{ground states}Calculated magnetic energies $E_{\mathrm{m}}$ (eV/fu) and magnetic moments $m_{\mathrm{Cr}}$ ($\mu_\mathrm{B}$/Cr) for the possible magnetic states in Sr$_2$Cr$_3$As$_2$O$_2$. See the text for the notations of N, F, A, C, G, S and K. K$_a$ (G$_c$) denotes that the moment is parallel to the $a$ ($c$) axis. For $E_{\mathrm{m}}$=0, it actually converges to nonmagnetic state.}
\begin{tabular}{c|c|c|c}
\hline
\hline
\quad Magnetic structure&$E_{\mathrm{m}}$& $m_{\mathrm{CrI}}$ &$m_{\mathrm{CrII}}$\\
\hline
Cr$^\mathrm{I}$N$-$Cr$^{\mathrm{II}}$N & 0 & 0&0 \\
\hline
Cr$^\mathrm{I}$N$-$Cr$^{\mathrm{II}}$C & $-$0.810 & 0&2.58 \\
Cr$^\mathrm{I}$N$-$Cr$^{\mathrm{II}}$F & $-$0.450 & 0&2.22 \\
Cr$^\mathrm{I}$N$-$Cr$^{\mathrm{II}}$S &0&0&0 \\
\hline
Cr$^\mathrm{I}$K$-$Cr$^{\mathrm{II}}$N & 0 & 0&0 \\
Cr$^\mathrm{I}$F$-$Cr$^{\mathrm{II}}$N & $-$0.967 & 3.11&0 \\
Cr$^\mathrm{I}$S$-$Cr$^{\mathrm{II}}$N & $-$1.069 & 3.05&0 \\
\hline
Cr$^\mathrm{I}$A$-$Cr$^{\mathrm{II}}$F  & $-$1.359 & 3.14&2.19 \\
Cr$^\mathrm{I}$F$-$Cr$^{\mathrm{II}}$F  & $-$1.303 & 3.14&2.2 \\
Cr$^\mathrm{I}$K$-$Cr$^{\mathrm{II}}$F  & $-$1.559 & 3.02&2.11 \\
Cr$^\mathrm{I}$S$-$Cr$^{\mathrm{II}}$F  & $-$1.459 & 3.07&2.13 \\
\hline
Cr$^\mathrm{I}$F$-$Cr$^{\mathrm{II}}$S  & $-$1.228 & 3.12&2.55 \\
Cr$^\mathrm{I}$K$-$Cr$^{\mathrm{II}}$S  & 0 & 0&0 \\
\hline
Cr$^\mathrm{I}$K$-$Cr$^{\mathrm{II}}$C & $-$1.967 & 3.10&2.60 \\
Cr$^\mathrm{I}$F$-$Cr$^{\mathrm{II}}$C & $-$1.790 & 3.09&2.59 \\
Cr$^\mathrm{I}$S$-$Cr$^{\mathrm{II}}$C & $-$1.880 & 3.03&2.59 \\
\hline
Cr$^\mathrm{I}$F$-$Cr$^{\mathrm{II}}$G & $-$1.772 & 3.08&2.65 \\
Cr$^\mathrm{I}$A$-$Cr$^{\mathrm{II}}$G &$-$1.755 & 3.09&2.57 \\
Cr$^\mathrm{I}$S$-$Cr$^{\mathrm{II}}$G & $-$1.871 & 3.03&2.57 \\
Cr$^\mathrm{I}$K$-$Cr$^{\mathrm{II}}$G& $-$1.973 & 2.98&2.59 \\
\hline
Cr$^\mathrm{I}$K$_{a}-$Cr$^{\mathrm{II}}$G$_c$ & $-$2.198 & 2.58&2.58 \\
Cr$^\mathrm{I}$K$_{a}-$Cr$^{\mathrm{II}}$G$_a$ & $-$2.208 & 2.97&2.59 \\
Cr$^\mathrm{I}$K$_{c}-$Cr$^{\mathrm{II}}$G$_c$ & $-$2.208 & 2.97&2.59 \\
\hline
\hline
\end{tabular}
\end{table}

We note that the same G-type magnetic order in the Cr$_2$As$_2$ layers was concluded in BaCr$_2$As$_2$\cite{singh}, BaFe$_{2-x}$Cr$_x$As$_2$ ($x>$ 0.3)\cite{sefat-Cr}, EuCr$_2$As$_2$\cite{hossain} and LaCrAsO\cite{park}, suggesting common magnetic exchange interactions in Cr$_2$As$_2$ layers. The dominant in-plane exchange couplings are usually modeled by the nearest-neighbor ($J_1$) and next-nearest-neighbor ($J_2$) interactions. According to the $J_{1}-J_{2}$ model\cite{johnston}, a 2D N\'{e}el AF configuration is stable if $J_{1}>0$ and $J_{1}>2J_{2}$. As stated previously, the decrease in the As$-$Cr$^{\mathrm{II}}-$As bond angle may reduce $J_{2}$ remarkably, which stabilizes the N\'{e}el AF state. As for the CrO$_2$ planes, the $J_{2}$ value should be negligibly small, resulting in a N\'{e}el AF state as obtained from the calculation above.

The magnetic moments mostly converge to 3.0(1) $\mu_\mathrm{B}$ and 2.60(5) $\mu_\mathrm{B}$ for Cr$^\mathrm{I}$ and Cr$^{\mathrm{II}}$, respectively. Both are less than the expected value of 4.0 $\mu_\mathrm{B}$ for the high-spin state of the isolated Cr$^{2+}$ ions. Similar reduced moments were also reported in BaCr$_2$As$_2$\cite{singh} and EuCr$_2$As$_2$\cite{hossain}, which is primarily due to the itinerancy of Cr-3$d$ electrons as well as the $d-p$ hybridizations. Owing to weaker $d-p$ hybridization in Cr$^\mathrm{I}$O$_2$ planes, the larger moment for Cr$^\mathrm{I}$ seems to be a reasonable result.

\begin{figure}
\includegraphics[width=8cm]{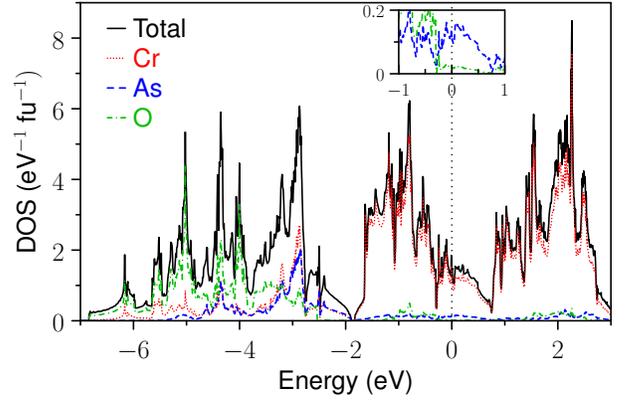}
\caption{\label{dos} (Color online) Total and projected electronic density of states (DOS) based on one spin for the G-type ground state of Sr$_2$Cr$_3$As$_2$O$_2$.}
\end{figure}

Figure~\ref{dos} shows the electronic DOS for the KG-type ground state of Sr$_2$Cr$_3$As$_2$O$_2$, from which we may have the following remarks. First of all, Sr atoms contributes vanishingly small weight in the energy window from $-7$ to +3 eV, consistent with the ionic core-shell state. Secondly, the oxygen 2$p$ states are almost fully occupied, and the DOS value is vanishingly small at $E_\mathrm{F}$, indicating very weak Cr3$d-$O2$p$ hybridization. Thirdly, for the As-4$p$ states, finite DOS at $E_\mathrm{F}$ (about 8.2\% of the total DOS) is observable, suggesting modest $d-p$ hybridizations between As-4$p$ and Cr-3$d$ orbitals. Finally, the DOS in the range of ($E_\mathrm{F}-$2) eV$<E<$ ($E_\mathrm{F}$+3) eV are mostly contributed from the Cr-3$d$ states. This is in contrast with the result of BaCr$_2$As$_2$ where the As-4$p$ orbitals comparably contribute the DOS near $E_\mathrm{F}$\cite{singh}. The dominant 3$d$ character around the $E_\mathrm{F}$ in Sr$_2$Cr$_3$As$_2$O$_2$ resembles the scenario in iron pnictides such as BaFe$_2$As$_2$\cite{singh2}, although the magnetic ground states are different.

The bare bandstructure DOS at $E_\mathrm{F}$ is calculated to be $N_{\mathrm{bs}}(E_{\mathrm{F}}$)=2.92 eV$^{-1}$fu$^{-1}$ based on both spins. This corresponds to a Sommerfeld coeficient, $\gamma_{0}=\frac{1}{3}\pi^{2}k_{\mathrm{B}}^2N(E_{\mathrm{F}})$=6.89 mJ K$^{-2}$ mol$^{-1}$. Thus, the mass renormalization factor is $\gamma/\gamma_{0}$=3.8, which is too high to be ascribed merely to electron-phonon interactions. This means that the electron-magnon couplings are likely to play a dominant role, as is suggested by the $\rho(T)$ behavior.

\begin{figure}
\includegraphics[width=8cm]{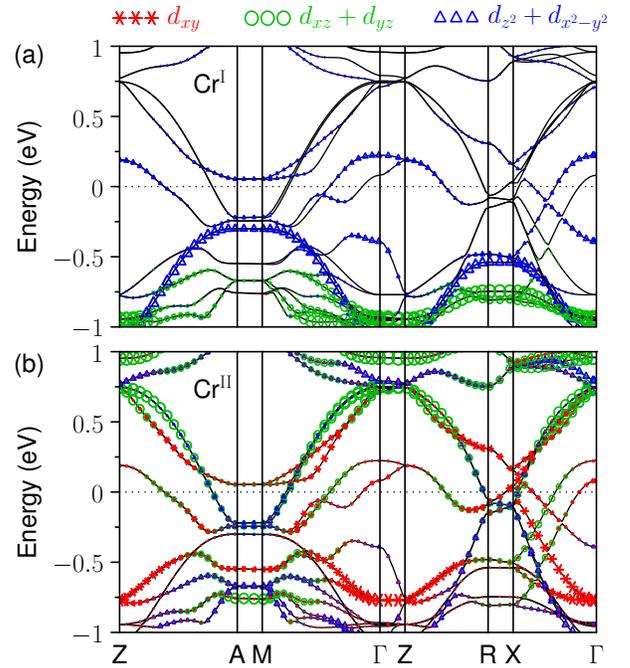}
\caption{\label{band} (Color online) Band structures of KG-type antiferromagnetic Sr$_2$Cr$_3$As$_2$O$_2$ near $E_\mathrm{F}$. The contributions from different 3$d$ orbitals in the Cr$^\mathrm{I}$ (a) and Cr$^\mathrm{II}$ (b) sites are distinguished by different symbols and colors. The symbols' size represents the relative weight.}
\end{figure}

\begin{figure}
\includegraphics[width=8cm]{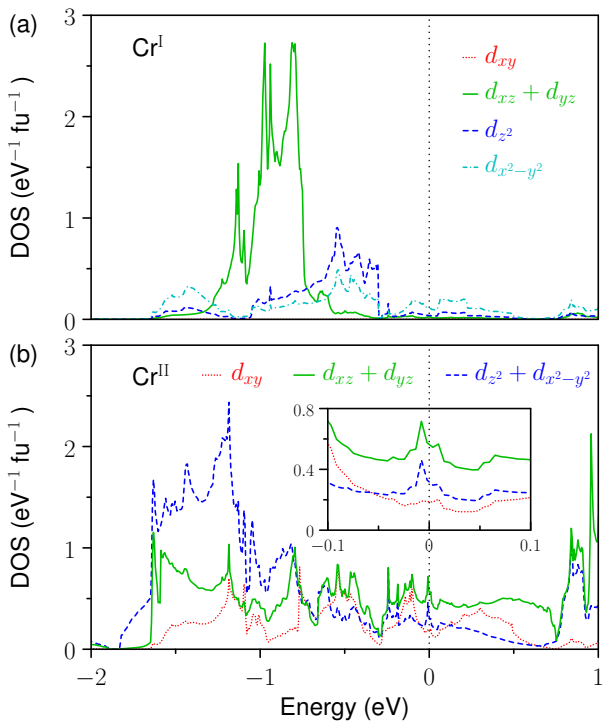}
\caption{\label{Cr-dos} (Color online) Projected density of states (DOS) to different groups of 3$d$ orbitals in the Cr$^\mathrm{I}$ (a) and Cr$^\mathrm{II}$ (b) sites in KG-type antiferromagnetic Sr$_2$Cr$_3$As$_2$O$_2$. Inset of (b) shows an enlarged plot near $E_\text{F}$.}
\end{figure}

\begin{figure}
\includegraphics[width=8cm]{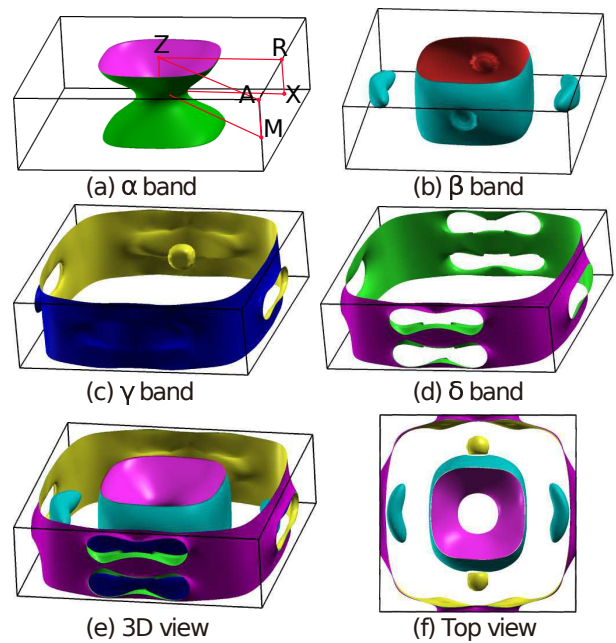}
\caption{\label{fs} (Color online) Calculated Fermi-surface sheets (FS) of the KG-type antiferromagnetic Sr$_2$Cr$_3$As$_2$O$_2$. (a)-(d): 3D views of $\alpha$-, $\beta$-, $\gamma$- and $\delta$-band FS. (e) and (f): the merged Fermi surface.}
\end{figure}

Now that the Cr-3$d$ states are predominant around the $E_\mathrm{F}$, we focus on the electronic band structures (Fig.~\ref{band}) and the corresponding DOS (Fig.~\ref{Cr-dos}), contributed from the Cr$^\mathrm{I}$ and Cr$^{\mathrm{II}}$ 3$d$ orbitals, respectively. With the tetragonal symmetry, the $d_{yz}$ and $d_{zx}$ bands are virtually degenerate. We also found that $d_{z^{2}}$ and $d_{x^{2}-y^{2}}$ are essentially (basically) degenerate for Cr$^{\mathrm{II}}$ (Cr$^{\mathrm{I}}$) 3$d$ orbitals. Hence the two groups of orbitals are put together. As may be seen, there are four energy bands cross the $E_\mathrm{F}$, consistent with the metallic behavior observed. No bands cross $E_\mathrm{F}$ along the $\Gamma-Z$ line, consistent with the 2D nature. Fig.~\ref{fs} shows the corresponding four Fermi-surface sheets (FS). Except for the $\alpha$-band FS, other three FS are basically cylindrical, in accordance with the flat bands along the $A-M$ line. Two pairs of small hole pockets can be seen in $\beta$ and $\gamma$ bands along \textbf{k}$_x$ and \textbf{k}$_y$ direction, respectively. In comparison, BaCr$_2$As$_2$ has three FS, one small rounded cube around $\Gamma$, and two cylinders centered at $\Gamma$ and along the \textbf{k}$_z$ direction\cite{singh}.

We note that the two Cr atoms contribute the band structures and DOS very differently. For the Cr$^\mathrm{I}$ atoms, the $d_{x^{2}-y^{2}}$ orbital contributes most of the hole states for the $\alpha$ band. This is related to the crystal-field effect, which particularly lifts the energy level of $d_{x^{2}-y^{2}}$ orbitals. For the Cr$^\mathrm{II}$ atoms, in contrast, every 3$d$ orbitals contribute the states around $E_\mathrm{F}$. The tetrahedron crystal field lifts the $t_2$ ($d_{xy}$, $d_{yz}$ and $d_{zx}$) levels, and among them, $d_{yz}$ and $d_{zx}$ are lifted further owing to the tetrahedron distortion. As a result, $d_{yz}$+$d_{zx}$ of Cr$^\mathrm{II}$ dominates the DOS at $E_\mathrm{F}$, resembling the scenario in iron pnictides\cite{2band}. Another interesting issue is that $E_\mathrm{F}$ locates nearby a van Hove singularity, which suggests the possibility of FM instability with a slight hole doping.

\subsection{\label{subsec:level6}On the origin of magnetic transitions}
With the above physical property measurements and first-principles calculations, we may further discuss the possible origin of the magnetic transitions in Sr$_2$Cr$_3$As$_2$O$_2$. Before the discussion, let us first review the magnetic transitions established in the closely related materials Sr$_2$Mn$_3$As$_2$O$_2$, Sr$_2$Mn$_3$Sb$_2$O$_2$ and LaCrAsO. The Mn moments in the Mn$_2$As$_2$ layers of Sr$_2$Mn$_3$As$_2$O$_2$ were found to order in a G-type antiferromagnetism below 340 K\cite{Mn2322B}. The ordered moment is 3.4 $\mu_\mathrm{B}$ at 4 K. Nevertheless, the $\chi(T)$ data between 80 K and 300 K are Curie-Weiss like\cite{Mn2322A}, which has to be due to the disordered Mn local moments in the MnO$_2$ planes. The susceptibility shows a hump-like feature at $\sim$60 K, corresponding to a 2D short-range magnetic correlations as revealed by the neutron diffractions\cite{Mn2322B}. For the analogous oxyantimide Sr$_2$Mn$_3$Sb$_2$O$_2$, the Mn spins in the MnO$_2$ planes order in K-type AFM configuration below 65 K with a significantly larger moment of 4.2 $\mu_\mathrm{B}$. For LaCrAsO\cite{park}, the magnetic susceptibility decreases with decreasing temperature below 500 K, suggesting 2D AFM correlations. Neutron diffractions reveal G-type AFM ordering at room temperature with a spin moment of 1.57 $\mu_\mathrm{B}$ (the moment was expected to be increased substantially at low temperatures) along the $c$ axis.

For Sr$_2$Cr$_3$As$_2$O$_2$, our theoretical calculations show that the Cr$_2$As$_2$ layers have a G-type magnetic ground state, similar to the cases in BaCr$_2$As$_2$\cite{singh}, EuCr$_2$As$_2$\cite{singh} and LaCrAsO\cite{park}. The linear $\chi(T)$ above 310 K in Sr$_2$Cr$_3$As$_2$O$_2$ also suggests 2D short-range spin correlations. Therefore, the AFM ordering at 291 K in Sr$_2$Cr$_3$As$_2$O$_2$ is likely associated with the AFM ordering in the Cr$_2$As$_2$ layers.

Next, the magnetism in the CrO$_2$ sheets is more 2D because the interplane magnetic coupling cancels out owing to the body-centered Cr$^{\mathrm{I}}$ sublattice. Thus, the magnetic transition temperature is expected to be relatively low, and the magnetism is basically 2D, like the case in Sr$_2$Mn$_3$As$_2$O$_2$. The observed broad transition around $T_{\mathrm{M3}}$ coincides with the expectation, which is then probably ascribed to the Cr$^{\mathrm{I}}$ short-range spin ordering/correlations. The absence of a long-range order, as predicted by our calculations above, may be due to the extreme two dimensionality (note that the possible interplane magnetic coupling between CrO$_2$ and Cr$_2$As$_2$ layers also cancels out owing to the geometry).

Finally, we make further remarks on the weak FM transition at $T_{\mathrm{M2}}$=230 K. Although our preliminary measurements and observations suggest an extrinsic nature for the phenomenon, the opposite conclusion cannot be fully ruled out. In case that the FM transition is intrinsic, here we propose two possible origins. One possibility is a local-moment ferromagnetism, arising from a slight spin-canting of the Cr moments. Another possibility is an itinerant ferromagnetism. In particular, the As-4$p$ electrons, with a small fraction near $E_\mathrm{F}$, could be spin-polarized. Such a nontrivial scenario was recently found and demonstrated in Ba$_{1-x}$K$_x$Mn$_2$As$_2$ which shows weak FM for $x>$0.19\cite{bjk,pandey,ueland}. Here the DOS of As-4$p$ also shows a maximum at $E_\mathrm{F}$ (the inset of Fig.~\ref{dos}), which could favor a FM instability.

\section{\label{sec:level4}Concluding Remarks}

We have synthesized a new layered chromium oxypnictide, Sr$_2$Cr$_3$As$_2$O$_2$. The crystal structure bears similarities with those of cuprate and iron-based superconductors, by possessing both CrO$_2$ planes and Cr$_2$As$_2$ layers. Magnetic transitions associated with the distinct Cr have been identified. Above room temperature, 2D AFM correlations in the Cr$_2$As$_2$ layers are suggested from the linear temperature dependence of magnetic susceptibility. The transition at 291 K is likely to be AFM with a G-type order. The broad transition around 38 K seems to occur in the CrO$_2$ planes where short-range AFM order is anticipated. Needless to say, one needs further experiments such as neutron diffractions to clarify the above preliminary conclusions.

Corresponding to the 2D-like crystal structure, the magnetism and the electronic structure also show 2D characteristic. The magnetic energy is dominated by the in-plane exchange couplings, and the magnetic ground state is theoretically predicted to be of a KG type order. The Fermi-surface sheets are basically 2D. Comparison of the bare bandstructure density of states at the Fermi level with the low-temperature specific-heat result yields a mass renormalization factor of $\sim$4, suggesting remarkable electron-magnon couplings. The electronic states around the Fermi energy are primarily contributed from the Cr-3$d$ orbitals, suggesting modest $d-p$ hybridizations. We also found that $d_{yz}$ and $d_{zx}$ dominate the DOS at $E_\mathrm{F}$. Based on these properties, we argue that the material could be a potential candidate for superconductivity.

\begin{acknowledgments}
We thank C. Cao for the help in the calculations. This work was supported by the National Basic Research Program of China (Grant No. 2011CBA00103), the National Science Foundation of China (Grant Nos. 11474252 and 11190023), and the Fundamental Research Funds for the Central Universities of China.
\end{acknowledgments}

\bibliography{2322}

\begin{thebibliography}{60}
\bibitem [{ein()}]{einstein}%
  \BibitemOpen
  \href@noop {} {}\bibinfo {note} {We also tried to use a single Debye term
  describing the phonon contribution, but the data fitting was not
  successful.}\BibitemShut {Stop}%
\bibitem [{s-t()}]{s-type}%
  \BibitemOpen
  \href@noop {} {}\bibinfo {note} {Since the magnetic energy is dominated by
  the in-plane exchange couplings, we first calculated the energy gain for a
  relatively simple magnetic structure, i.e. in-plane striped AF with
  interplane F here. Te result shows that such a in-plane magnetic order has
  the magnetic energy $\sim$0.53 eV higher than that of the N\'{e}el AF order.
  Thus, the in-plane striped AF state with interplane AF does not need to be
  calculated anymore.}\BibitemShut {Stop}%
\end{thebibliography}

\end{document}